# Negative bulk modulus and possibility of loss of elastic stability near tricritical transitions in thin films on substrates

A. P. Levanyuk [1], S. A. Minyukov [2], I. B. Misirlioglu [3]

[1] *University of Washington, Department of Physics, Seattle, WA 98105 USA*
[2] *Shubnikov Institute of Crystallography of Federal Scientific Research Centre "Crystallography and Photonics" of Russian Academy of Sciences, Moscow 119333, Russia*
[3] *Faculty of Engineering and Natural Sciences, Sabanci University, Tuzla/Orhanli 34956 Istanbul, Turkey*

Within the Landau-like approach we study anomalies of elastic moduli at phase transitions in thin films on substrates. We consider the case where, similar to many experimental cases, the first-order transition in free crystal would convert into a second order in the film if the system remained homogeneous. It is shown, however, that apart from its questionable thermodynamic profitability, the homogeneous state of low-symmetry phase may become absolutely unstable which is signaled by changing of sign of its bulk modulus.

---

Author to whom correspondence should be addressed: A. P. Levanyuk, levanyuk@u.washington.edu

# I. Introduction

Ferroelectric thin films on substrates are actively studied during last two decades. At the same time properties of these films near phase transition from paraelectric to ferroelectric phase are not properly understood. This is unfortunate because this phase transition is considered since long ago as a key point to understand properties of ferroelectrics (see, e.g., [1]). One of the experimental observations resisting reliable explanation is that continuous, seemingly second order transitions are much "worse" in thin films and other nanosystems than in bulk crystal: the anomalies are smeared, slow relaxations are frequently observed. These specific features persist even in very high quality systems and that is why usual reference to structural imperfections does not seem sufficient. To look for other explanations it is useful to emphasize that a more adequate name of what is often called second order phase transitions in thin films on substrates (e. g. paraelectric-ferroelectric phase transition in $BaTiO_3$ or $PbTiO_3$ films on a $SrTiO_3$ substrate) are first order transitions in partially constrained systems. Indeed, in bulk $BaTiO_3$ and $PbTiO_3$ the ferroelectric phase transition is of first order and what happens because of clamping due to substrate is not immediately clear. Theoretical conclusion that the transition in a fully clamped $BaTiO_3$ would be of second order comes back to Devonshire [2] who considered homogeneous states only. It was found that this conclusion holds also for partially clamped $BaTiO_3$ (as well for $PbTiO_3$) films the clamping being due to the substrate [3] and the assumption about the homogeneity is once more adopted. At the same time it was known since long ago that first order phase transitions in constrained or partially constrained systems occur often through inhomogeneous states where two phase regions form. This makes the above mentioned assumption about homogeneity of the system at least questionable [4]. In what follows in the rest of the paper, when mentioning tricritical or second order transition in thin films on substrates, we do not mean phase transitions that actually are expected to occur but "virtual" transitions obtained theoretically under the assumption of homogeneity.

The present paper is aimed to contribute to making the situation more clear.. Let us explain first of all why it is not clear until now. It was already mentioned that formation of two-phase regions at the phase transition is known since long ago. The simplest example is formation of meniscus at cooling of an ampule with water vapor, i.e. formation of two-phase state of water. In solids, any inhomogeneity in mass density is accompanied by shear stresses both in the transforming material and in the surrounding medium which makes the phenomenon far from being simple. For crystalline films on substrates, formation of transversely modulated structures consisting of regions of the two phases has been predicted by Roytburd [5]. Parameters of such a structure has been discussed in [6] for the case of sufficiently thick films. For the opposite case of extremely thin films one can seemingly apply arguments of [7] where first order surface phase transitions were considered and formation of two-phase periodic structures has been also predicted. But in the latter paper it has been virtually supposed that the first order transition could take place also homogeneously in the constrained layer, e.g. in an academic case of infinite surface energy of the interphase boundaries, rendering a two phase state impossible. This is, however, not our case, where, as it was mentioned, a first order phase transition in a free crystal converts into a second order one if it would occur homogeneously [2], [3]. The arguments of Roytburd are valid for any first order transition but in sufficiently thick films. At the same time it is very thin films which are of the main interest now. It is important, of course, to reveal if the modulated structures predicted by Roytburd remain in some form in films of any thickness on substrates or if they disappear at what film thickness this occurs. But these questions are not treated in this paper.

Let us assume that the two-phase state indeed disappears at some thickness and consider smaller thicknesses than this one. Or assume that it remains but the period of the modulated structure becomes very large at small thicknesses similar to elastic domain structure considered in [8]. In this case we can seemingly have homogeneous state in very thin films with finite (though, may be, quite realizable) lateral dimensions. Or, recalling that nucleation is needed to form a two-phase state, we imagine that nucleation centers are absent and we have homogeneous state which is metastable but quite observable. In other words, it may seem that we can imagine realistic situations where, because of avoiding somehow two-phase states, we obtain homogeneous states in the clamped system where the conclusions of [2], [3] are applicable. In this paper we show that if the virtual homogeneous transition assumed by [2], [3] is close to tricritical point this may be impossible. The homogeneous state may lose its elastic stability before reaching the phase transition when the system is heated from low temperatures. We estimate that width of the region where the homogeneous state is unstable may be of several degrees so that it treats about potentially observable phenomenon. Even before reaching the instability point the bulk elastic modulus of material of the film becomes negative which also might be of interest..

Specifically, we shall show that, due to a peculiarity of phase transitions in thin films on substrates, the positive definiteness of the quadratic form presenting the elastic energy can be violated due to anomalies in the elastic moduli. In a non-restricted elastic body, this would be considered as leading to loss of elastic stability. For a restricted, i. e. partially or completely clamped body, an additional analysis is needed, which is mainly beyond the scope of this paper and, furthermore, this analysis can be rather different for differently restricted systems. Thus, in this paper we intend rather to formulate a problem than to solve it.

A consequence of loss of elastic stability of the homogeneous state in a thin film on a substrate is, most probably, formation of an inhomogeneous state. The character of this state is unknown at the moment. It might be identical to the two-phase states mentioned above and in this case the elastic instability is the spinodal point for two-phase state formation (never mentioned before). But it might be also a completely new state competing with the above described two-phase state. Only further studies can make it clear which of the two options is realized.

It has been already mentioned that it is due to a peculiarity of phase transitions in films on substrates that the anomalies of elastic moduli may lead to loss of positive definiteness of elastic energy. This peculiarity consists in constancy, at the phase transition, of some components of the strain tensor fixed by the substrate. This is unlike majority of "text-book" transitions which occur at constant pressure. To treat the elastic anomalies at phase transitions in films on substrates it is convenient to explicitly take into account the strains from the very beginning. We illustrate this method in Sec. II where we apply it to a problem which was solved long ago [9] by another method, more convenient for this particular problem but inapplicable to phase transitions in films on substrates. It treats about anomaly of bulk elastic modulus at tricritical phase transition in a free isotropic body. The tricritical transition is mainly considered also for films on substrates because the temperature interval where the elastic energy is not positive definite is the most broad just at such a transition. It becomes narrower and finally disappears at moving off from the tricritical point. Recall that the bulk modulus goes to zero at tricritical transition in a free body and since positiveness of the bulk modulus is the condition of elastic stability in unrestricted systems, one can say that the system is driven to the boundary of elastic stability at this transition but does not cross the boundary. After reproducing the old result we consider once more tricritical transition but this time in a film on a substrate (Sec. III). Similar to the preceding case there is also an elastic modulus which goes to zero at the phase transition but this time it is not the bulk but longitudinal modulus. Since this modulus is larger than the bulk one the latter changes its sign and the elastic energy stops to be positive definite well before the transition. Unlike the previous case the boundary of stability may be crossed. In the proceeding section (Sec. IV) the real symmetry of BaTiO$_3$ and PbTiO$_3$ on a (001) cubic substrate is taken into account assuming that it is a tricritical phase transition in the thin film. It is also assumed that the misfit strains which convert the cubic paraelectric phase into a tetragonal one impose the polar axis to be perpendicular to the film plane, i.e. there is a one-component ferroelectric order parameter as it takes place, e. g., for BaTiO$_3$ on SrTiO$_3$. Loss of positive definiteness of the elastic energy is demonstrated once more and numerical values of material parameters of these crystals are used to estimate the temperature width of the region where the elastic energy is not positive definite. Ideal metallic electrodes are assumed to exclude the effects of depolarizing field which are superposed with the elastic phenomena otherwise. In Sec. V we give elementary illustrations of difference of the elastic stability conditions in restricted and in unrestricted systems. In Sec. VI further possible studies of the problem outlined in this paper are discussed.

## II. Tricritical transition in isotropic medium

As far as the effects of depolarizing field are not taken into account our consideration is relevant also for non-ferroelectric phase transitions and we designate the order parameter as $\eta$. The Landau-like free energy has the form

$$F = \frac{\alpha}{2}\eta^2 + \frac{\beta}{4}\eta^4 + \frac{\gamma}{6}\eta^6 + r\eta^2(u_{11} + u_{22} + u_{33}) + \frac{\lambda + 2\mu}{2}(u_{11}^2 + u_{22}^2 + u_{33}^2) + \lambda(u_{11}u_{22} + u_{11}u_{33} + u_{22}u_{33}) + 2\mu(u_{12}^2 + u_{13}^2 + u_{23}^2),$$ (1)

where $\lambda$ and $\mu$ are, correspondingly, the Lame coefficient [10] and the shear modulus of the high-symmetry phase. This phase in absence of stresses and with temperature expansion neglected is considered as the reference state for the strains. Differentiating $F$ with respect to $u_{11}$ one obtains

$$\sigma_{11} = (\lambda + 2\mu)u_{11} + \lambda(u_{22} + u_{33}) + r\eta^2 = \lambda u_{ii} + 2\mu u_{11} + r\eta^2,$$ (2)

where $u_{ii} = u_{11} + u_{22} + u_{33}$. Equations for $\sigma_{22}$ and $\sigma_{33}$ are analogous. The $\sigma_{12}$ and other non-diagonal components of $\sigma_{ik}$ tensor are of no interest for what follows. For the order parameter in the low-symmetry phase we have:

$$\alpha + \beta\eta^2 + \gamma\eta^4 + 2ru_{ii} = 0$$ (3)

Consider first a free crystal, i. e. $\sigma_{ik} = 0$ for any $i, k$. One finds that the non-diagonal components of $u_{ik}$ are zero and from three equations of type of Eq. 2:

$$u_{11} = u_{22} = u_{33} = -\frac{r\eta^2}{3\lambda + 2\mu} = -\frac{r\eta^2}{3K}, \tag{4}$$

where $K$ is the bulk modulus. Substituting this into Eq. 3 we obtain:

$$\alpha + (\beta - 2r^2/K)\eta^2 + \gamma\eta^4 = 0 \tag{5}$$

For simplicity we assume that the only coefficient which depends on temperature is $\alpha$. If the phase transition in the free crystal is of the first order then $\beta < 2r^2/K$. For further discussion it makes sense to find the elastic moduli of the low-symmetry phase. We shall consider the case where $\beta = 2r^2/K$, i. e. the phase transition corresponds to tricritical point. Then for values of the order parameter and the strain components in free crystal of the low-symmetry phase, one has:

$$\eta_f^2 = \sqrt{-\alpha/\gamma}, u_{11f} = u_{22f} = u_{33f} = -r\eta_f^2/3K \tag{6}$$

If, e. g., a stress $\sigma_{11}$ is applied to a formerly free crystal, the value of $\eta^2$ changes to $\eta^2 = \eta_f^2 + \delta\eta^2$ as well as the values of strains: $u_{11} = u_{11f} + \delta u_{11}$ etc. where $\delta\eta^2$ and $\delta u_{11}$ are changes in $\eta^2$ and $u_{11}$ due to the stress. Assuming that the changes are small since we are interested in linear response we have from Eq. 3:

$$(\beta + 2\gamma\eta_f^2)\delta\eta^2 = -2r\delta u_{ii} \tag{7}$$

and Eq. 2 acquires the form

$$\sigma_{11} = \lambda\delta u_{ii} + 2\mu\delta u_{11} + r\delta\eta^2 = \left(\lambda - \frac{2r^2}{\beta + 2\gamma\eta_f^2}\right)\delta u_{ii} + 2\mu\delta u_{11} \tag{8}$$

From Eq. 8 and taking into account that $\beta = 2r^2/K$, we find that the Lame coefficient of the low-symmetry phase:

$$\tilde{\lambda} = \lambda - \frac{K}{1 + (K/r^2)\sqrt{-\alpha\gamma}} \tag{9}$$

becomes temperature-dependent and decreases as the phase transition is approached. The bulk modulus of the low-symmetry phase is

$$\tilde{K} = \tilde{\lambda} + \frac{2\mu}{3} = K\frac{(K/r^2)\sqrt{-\alpha\gamma}}{1 + (K/r^2)\sqrt{-\alpha\gamma}} \cong \frac{K^2}{r^2}\sqrt{-\alpha\gamma}, \tag{10}$$

where the approximate equality holds when one is sufficiently close to the tricritical point. We see that the bulk modulus is zero at the tricritical point (Fig. 1). This is a very old result obtained by another method by Landau in 1934 [9].

## III. Tricritical transition in elastically isotropic thin film on substrate

Suppose now that we have an epitaxial film on an isotropic substrate whose plane contains axes $1$ and $2$. We suppose that the substrate is of infinite thickness so that there is no homogeneous strain in the substrate and only the film is strained because of the misfit. When defining the strains we shall consider the film in the symmetrical phase as the reference state, i. e. $u_{11} = u_{22} = u_{33} = 0$ in this phase. Because of perfect clamping of the film due to the substrate the homogeneous strains $u_{11}, u_{22}$ are also zero in the low-symmetry phase while the homogeneous $u_{33}$ changes due to the stress free surface ($\sigma_{33} = 0$). The stresses in the film change also and when writing $\sigma_{ik}$ we shall mean not the whole stresses but only those occurring on top of the misfit-induced stresses in the symmetrical phase. Considering homogeneous states only one can calculate $u_{33}$ from the condition of the free surface:

$$\sigma_{33} = (\lambda + 2\mu)u_{33} + r\eta^2 = 0 \tag{11}$$

One can, of course, calculate also $\sigma_{11}$ and $\sigma_{22}$ but they are of no interest for what follows. Substituting $u_{33} = -r\eta^2/(\lambda + 2\mu)$, $u_{11} = u_{22} = 0$ into Eq. 3 one obtains:

$$\alpha + (\beta - 2r^2/(\lambda + 2\mu))\eta^2 + \gamma\eta^4 = 0 \tag{12}$$

If the phase transition in the film is of second order then $\beta > 2r^2/(\lambda + 2\mu) = 2r^2/\overline{K}$, where $\overline{K}$ is the longitudinal modulus of symmetrical phase. Since $\overline{K} > K$ it is quite possible that $2r^2/K > \beta > 2r^2/\overline{K}$, i. e. the phase transition which is of first order in a free crystal becomes of second order in the film on a substrate (if it remains homogeneous) which was observed in [3] for $BaTiO_3$ and $PbTiO_3$. To calculate the Lame coefficient in the low-symmetry phase we should once more find the dependence of $\eta^2$ on strains. Note that we are interested in strains and stresses which are local so that all the components of the strain tensor are allowed. Eq. 7 remains valid and in Eq. 8 we only have to substitute $\sigma_{11}$ by $\delta\sigma_{11}$ and $\sigma_{22}$ by $\delta\sigma_{22}$ ("test stresses") given that non-zero stresses $\sigma_{11}$ and $\sigma_{22}$ exist in the homogeneous state of low-symmetry phase starting from the phase transition onwards. Also, we should replace $\eta_f^2$ by $\eta_s^2$ where the subscript $s$ stands for "substrate". Once more we shall consider the tricritical transition, this time in the film, i. e. we put $\beta = 2r^2/\overline{K}$. Then we obtain

$$\widetilde{\lambda} = \lambda - \frac{\overline{K}}{1 + (\overline{K}/r^2)\sqrt{-\alpha\gamma}} \tag{13}$$

$$\widetilde{\widetilde{K}} = \widetilde{\lambda} + 2\mu = \overline{K}\,\frac{(\overline{K}/r^2)\sqrt{-\alpha\gamma}}{1+(\overline{K}/r^2)\sqrt{-\alpha\gamma}} \cong \frac{\overline{K}^2}{r^2}\sqrt{-\alpha\gamma} \qquad (14)$$

cf. Eqs. 9, 10. For the bulk modulus of low-symmetry phase we have now

$$\widetilde{K} = \widetilde{\lambda} + \frac{2\mu}{3} = \frac{K-\overline{K}+(K\overline{K}/r^2)\sqrt{-\alpha\gamma}}{1+(\overline{K}/r^2)\sqrt{-\alpha\gamma}} = \frac{-4\mu/3+(K\overline{K}/r^2)\sqrt{-\alpha\gamma}}{1+(\overline{K}/r^2)\sqrt{-\alpha\gamma}} \cong \frac{-4\mu}{3} + \frac{K\overline{K}}{r^2}\sqrt{-\alpha\gamma} \qquad (15)$$

We see that now that the longitudinal modulus is zero but the bulk modulus is negative ($\widetilde{K} = -4\mu/3$) at the transition. This means that the elastic energy in the film is not positive definite at the phase transition and, in fact, loss of the positive definiteness occurs well before the transition. (Fig. 2). To estimate the possible importance of this finding for real materials we consider in the next Section a virtual tricritical phase transition in cubic films on (001) cubic substrates using some of BaTiO$_3$ and PbTiO$_3$ parameters which allows us to make numerical estimates for a system, which is close to real ones.

## IV. Tricritical transition in cubic crystalline films

As it has already been mentioned above, the order parameter will be identified with the polarization component perpendicular to the film plane ($\eta = P_3$). Since the misfit strain is usually fairly small ( 2.2% for BaTiO$_3$ and 1.2% for PbTiO$_3$ on SrTiO$_3$) the influence of the misfit-induced tetragonality on the elastic moduli of the paraelectric phase can be neglected and what is usually called the Landau-Ginzburg-Devonshire free energy has the form:

$$F = \frac{\alpha}{2}\eta^2 + \frac{\beta}{4}\eta^4 + \frac{\gamma}{6}\eta^6 + r_1\eta^2 u_{33} + r_2\eta^2(u_{11}+u_{22}) + \\ \frac{1}{2}c_{11}(u_{11}^2+u_{22}^2+u_{33}^2) + c_{12}(u_{11}u_{22}+u_{11}u_{33}+u_{22}u_{33}) + 2\mu(u_{12}^2+u_{13}^2+u_{23}^2) \qquad (16)$$

The last term is not important for the rest of the Section and shall not be used below since the uniaxial polarization is coupled to diagonal strain components only. The governing equations for the ferroelectric state are:

$$\sigma_{11} = c_{11}u_{11} + c_{12}(u_{22}+u_{33}) + r_2\eta^2, \qquad (17)$$

$$\sigma_{22} = c_{11}u_{22} + c_{12}(u_{11}+u_{33}) + r_2\eta^2 \qquad (18)$$

$$\sigma_{33} = c_{11}u_{33} + c_{12}(u_{11}+u_{22}) + r_1\eta^2 \qquad (19)$$

$$\alpha + \beta\eta^2 + \gamma\eta^4 + 2r_1 u_{33} + 2r_2(u_{11} + u_{22}) = 0 \qquad (20)$$

In homogeneous state and we have once more $u_{11} = u_{22} = 0$ due to clamping and $\sigma_{33} = 0$ (stress free surface) so that

$$u_{33} = -r_1\eta^2 / c_{11} \qquad (21)$$

and value of the order parameter in homogeneous state is defined by the equation

$$\alpha + (\beta - 2r_1^2/c_{11})\eta^2 + \gamma\eta^4 = 0 \qquad (22)$$

from which one can obtain for $\eta$ in homogeneous state ($\eta_h$)

$$\eta_h^2 = \frac{-\hat{\beta} + \sqrt{\hat{\beta}^2 - 4\alpha\gamma}}{2\gamma}, \qquad (23)$$

where $\hat{\beta} = \beta - 2r_1^2/c_{11}$. To calculate the moduli in the low-symmetry phase we have now instead of Eq. 7:

$$(\beta + 2\gamma\eta_h^2)\delta\eta^2 = \left(2r_1^2/c_{11} + \sqrt{\hat{\beta}^2 - 4\alpha\gamma}\right)\delta\eta^2 = -2r_1\delta u_{33} - 2r_2(u_{11} + u_{22}). \qquad (24)$$

Similar to the isotropic case we obtain

$$\delta\sigma_{11} = \tilde{c}_{11}u_{11} + \tilde{c}_{12}u_{22} + \tilde{c}_{13}\delta u_{33}, \qquad (25)$$

$$\delta\sigma_{22} = \tilde{c}_{11}u_{22} + \tilde{c}_{12}u_{11} + \tilde{c}_{13}\delta u_{33}, \qquad (26)$$

$$\sigma_{33} = \tilde{c}_{33}\delta u_{33} + \tilde{c}_{13}(u_{11} + u_{22}), \qquad (27)$$

where

$$\tilde{c}_{11} = c_{11} - 2r_2^2 / \left(2r_1^2/c_{11} + \sqrt{\hat{\beta}^2 - 4\alpha\gamma}\right), \qquad (28)$$

$$\tilde{c}_{12} = c_{12} - 2r_2^2 / \left(2r_1^2/c_{11} + \sqrt{\hat{\beta}^2 - 4\alpha\gamma}\right), \qquad (29)$$

$$\tilde{c}_{13} = c_{12} - 2r_2 r_1 / \left(2r_1^2/c_{11} + \sqrt{\hat{\beta}^2 - 4\alpha\gamma}\right), \qquad (30)$$

$$\widetilde{c}_{33} = c_{11} - 2r_1^2 / \left(2r_1^2 / c_{11} + \sqrt{\widehat{\beta}^2 - 4\alpha\gamma}\right). \tag{31}$$

One sees that despite our simplification of the elastic energy of the paraelectric phase, the set of elastic moduli of the ferroelectric phase becomes more similar to that of a tetragonal crystal.

According to Sylvester's criterion, the conditions for the elastic energy to be positive definite as a function of $u_{11}$, $u_{22}$, $u_{33}$ are:

$$\widetilde{c}_{33} > 0, \widetilde{c}_{11} > |\widetilde{c}_{12}|, \widetilde{c}_{33} > 2\widetilde{c}_{13}^2 / (\widetilde{c}_{11} + \widetilde{c}_{12}). \tag{32}$$

For the sake of illustration consider tricritical transition ($\widehat{\beta} = 0$) and assume that $r_2 = 0$. The latter is fairly similar to BaTiO$_3$ where, according to some authors, see e. g. [11], one has $r_2/r_1 = 5 \times 10^{-2}$. Then we have:

$$\widetilde{c}_{11} = c_{11}, \widetilde{c}_{12} = \widetilde{c}_{13} = c_{12} \tag{33}$$

and

$$\widetilde{c}_{33} = c_{11} - 2r_1^2 / \left(2r_1^2 / c_{11} + \sqrt{-4\alpha\gamma}\right) \tag{34}$$

Since $\widetilde{c}_{11}$ and $\widetilde{c}_{12}$ are no more dependent on temperature and the first inequality in Eq. 32 is surely satisfied when the third one is satisfied we have to take into account the latter inequality only. The condition of loss of the positive definiteness reads:

$$c_{11} - 2r_1^2 / \left(2r_1^2 / c_{11} + \sqrt{-4\alpha\gamma}\right) = 2c_{12}^2 / (c_{11} + c_{12}) \tag{35}$$

or writing for $\alpha$, at

$$-\alpha = \frac{r_1^4}{\gamma c_{11}^2} \left(\frac{2\xi^2}{1 + \xi - 2\xi^2}\right)^2 \cong \frac{r_1^4}{\gamma c_{11}^2}, \tag{36}$$

where $\xi = c_{12}/c_{11}$ and it is taken into account that for BaTiO$_3$ the expression in the parentheses is about unity [12]. Using the coefficient $\alpha$ for BaTiO$_3$ [12] we find that the positive definiteness of elastic energy is lost at

$$T_{tc} - T \cong 15°C \tag{37}$$

Though no experimental example of exactly tricritical transition in thin films on substrates is known, the region of lack of the positive definiteness of elastic energy proves to be fairly wide and it might be sufficiently wide to reveal it experimentally also for second order transitions close to tricritical point.

## V. Second order phase transition in crystalline films

If the second order phase transition does not coincide with tricritical point the width of region where the elastic energy is not positive definite will be less, of course. To have an idea about how this interval diminishes with $\hat{\beta}$ attaining larger values it makes sense to consider once more the case $r_2 = 0$ abandoning the assumption that $\hat{\beta} = 0$. Then for condition of the positive definiteness loss we have instead of Eq. 35:

$$c_{11} - 2r_1^2 \Big/ \left( 2r_1^2/c_{11} + \sqrt{\hat{\beta}^2 - 4\alpha\gamma} \right) = 2c_{12}^2/(c_{11} + c_{12}). \tag{38}$$

Let us find the maximal value of $\hat{\beta}$ for which the loss of the positive definiteness still takes place, i.e. for which this loss coincides with the phase transition. Putting $\alpha = 0$ in Eq. 38 we obtain:

$$\hat{\beta} = 2\frac{r_1^2}{c_{11}}\frac{2\xi^2}{1+\xi-2\xi^2} \cong 2\frac{r_1^2}{c_{11}} \tag{39}$$

i. e. the loss occurs before the phase transition if

$$\beta < \frac{4r_1^2}{c_{11}} = 29.2 \times 10^8 \, Jm^5 C^{-4}, \tag{40}$$

where the numerical values of the constants of BaTiO$_3$ reported in [12] have been used. It is difficult to conclude if this condition is fulfilled for BaTiO$_3$ since values of $\beta$ are different according to different authors. It seems that the latest discussion of this question is given in [13] where the values of $\beta$ according to different authors can be found. The authors themselves propose for a free crystal that $\beta_{free} = -7.3 \times 10^9 + 16 \times 10^6 T \, Jm^5 C^{-4}$. We calculate $\beta$ for a clamped crystal using the value $\beta_{free}$ given in [13]. This calculation has been made several times beginning with Devonshire [2] (see also [11], [14]). In the notations of this paper it reads:

$$\beta_{free} = \beta - \frac{2}{3}\left( \frac{(r_1 + 2r_2)^2}{c_{11} + 2c_{12}} + \frac{(r_1 - r_2)^2}{c_{11} - c_{12}} \right). \tag{41}$$

Using the values of the constants of [12] one obtains:

$$\beta = -5.56 \times 10^9 + 16 \times 10^6 T \, Jm^5 C^{-4} \tag{42}$$

From Eqs. 42 and 40 one finds that if the phase transition occurs at $383K < T < 575K$ it is accompanied by the loss of positive definiteness of the elastic energy For another set of the material constants of BaTiO$_3$ [15] $\beta_{free} = -8.4 \times 10^8 \, Jm^5 C^{-4}$ and does not depend on temperature. Finding $\beta = 9 \times 10^8 \, Jm^5 C^{-4}$ we see that the inequality 40 is always fulfilled, i.e. the loss of positive definiteness of the elastic energy occurs for any temperature of the phase transition. In the third set of the constants [16] $\beta_{free} = -1.34 \times 10^6 (T - 381K) Jm^5 C^{-4}$, i. e. $\beta = 1.23 \times 10^9 - 1.34 \times 10^6 T$ the loss takes place also at any temperature of the phase transition.

## VI. Lack of positive definiteness of elastic energy and instability

Positive definiteness of the elastic energy is usually mentioned in books as the condition which puts restrictions on possible values of elastic constants of a material. The argument is that if this energy is not positive definite, the system can lower its energy by developing strain without any stress, i. e. the reference state for the strains is not stable. It is tacitly assumed that any strain is possible in the system. This is true, of course, for unrestricted systems but not necessarily for thin films on substrates which belong to partially restricted systems. In this case the analysis should be performed anew. To illustrate this let us consider stability with respect to strains $u_{11}$ and $u_{33}$ only supposing that all other strains are somehow prohibited (a very severe restriction). The body is supposed to be made from isotropic elastic material. For stability of the system we should demand the quadratic form in the variables $u_{11}$, $u_{33}$ :

$$\frac{\lambda + 2\mu}{2}\left(u_{11}^2 + u_{33}^2\right) + \lambda u_{11} u_{33} \tag{43}$$

to be positive definite. The conditions of positive definiteness of this quadratic form are:

$$\lambda + 2\mu, \ \mu(\lambda + \mu) > 0 \tag{44}$$

or, supposing the positiveness of $\mu$,

$$\lambda + \mu > 0 \tag{45}$$

or

$$K + \frac{\mu}{3} > 0. \tag{46}$$

If we allow also $u_{22}$, i.e. the quadratic form in question is

$$\frac{\lambda + 2\mu}{2}\left(u_{11}^2 + u_{33}^2 + u_{22}^2\right) + \lambda\left(u_{11} u_{33} + u_{11} u_{22} + u_{22} u_{33}\right) \tag{47}$$

and a third condition should be added to Eq.42 according to Sylvester's criterion

$$\lambda + \frac{2\mu}{3} > 0. \tag{48}$$

or

$$K > 0 \qquad (49)$$

Now the two other conditions are automatically satisfied together with this condition if $\mu > 0$. Comparing with Eq. 46 we see that now there is less space for elastic stability than in the previous case where for

$$0 > K > -\frac{\mu}{3} \qquad (50)$$

the system remains elastically stable despite $K < 0$. Of course, the both above examples are of pure illustrative character. In real systems it is not possible to impose such severe restrictions on the strain components. For films on substrates, for example, it is homogeneous strains $u_{11}$ and $u_{22}$ which are surely prohibited while other restrictions come from the boundary conditions at the film-substrate interface and reveal themselves in the process of solution of the full elasticity problem, which considers both homogeneous and inhomogeneous strains. This implies a far more involved but still quite doable analysis which is beyond the scope of this paper. Preliminary calculations for a slab of isotropic material with free upper and low surfaces show that Eq. 45 gives the correct condition of elastic stability of this system. Since $\tilde{\lambda} + 2\mu = 0$ at the tricritical transition temperature, the interval where not only positive definiteness of elastic energy but also the elastic stability is lost correspond to about half of the interval of negative bulk modulus (Fig. 3). In this interval we have**:**

$$-\mu > \tilde{\lambda} > -2\mu \qquad (51)$$

or

$$-\frac{\mu}{3} > \tilde{K} > -\frac{4\mu}{3} \qquad (52)$$

In a film on an infinitely rigid substrate the loss of stability also occurs but at a lower temperature. As we have already mentioned we do not know what the loss of elastic stability leads to. This also has to be studied.

## VII. Conclusion

We have shown that theoretical results for anomalies of elastic moduli near phase transitions in films on substrates may be fairly peculiar if the system is assumed to be homogeneous and the considered phase transition is near the tricritical point. The bulk modulus may become negative before the transition and the question about elastic stability of the system naturally arises. These peculiarities are due to that the constancy of certain strains is maintained at the transitions while the "classical" situation is when constancy of the stresses (or pressure) takes place. Note that if a phase transition is tricritical in a film on a substrate, it is of first order in a free sample with the same Curie temperature. As aforementioned in the introduction, a tricritical transition in a constrained system considered here is a theoretical expectation based on a simplifying assumption of behavior under constraints of materials with first order transitions. Specifically it is assumed that the system remains homogeneous under constraints throughout the entire temperature range. In this paper we have shown that this expectation might be wrong and the assumption about homogeneity invalid since the elastic stability of homogeneous state might be lost before the transition. How this loss of elastic stability can be observed is another question because in real life a two-phase state may arise before the stability loss as the two-phase state may become thermodynamically more profitable than the homogeneous one irrespective of the stability aspect. This may make observation of the stability loss not an easy task. One has to somehow avoid or hamper nucleation of the new phase. This is similar to the difficulty of observing spinodal points of many phase transitions. One may hope, however, that study of very thin films on substrates may be promising in this aspect because the thinner is the film the less is "the driving force" for formation of two-phase state according to the Roytburd's mechanism while the condition of loss of elastic stability remains intact.

**Acknowledgement**

# Figure Captions

**Figure 1.** Plot of the bulk modulus near tricritical transition in a stress-free isotropic body. The axes are normalized with respect to $\lambda$ and the temperature of the tricritical transition, $T_{tc}$, $\lambda = 2.5\mu$. Vertical dashed line is to indicate $T_{tc}$. We do not consider the temperature dependence of the modulus in the HSP region and assume it is constant. LSP: Low symmetry phase, HSP: High symmetry phase.

**Figure 2.** Plot of the normalized bulk modulus (red line) and the longitudinal modulus (blue line) of the isotropic film laterally restricted near the tricritical transition under the assumption of the homogeneous state. The axes are normalized with respect to $\lambda$ and the temperature of the tricritical transition, $T_{tc}$, $\lambda = 2.5\mu$. Vertical dashed line is to indicate $T_{tc}$. The shaded region is the range of temperature where bulk modulus of the film is negative. We do not consider the temperature dependence of the moduli in the HSP region and assume it is constant. LSP: Low symmetry phase, HSP: High symmetry phase.

**Figure 3.** Plot of the normalized bulk modulus (red line) and the longitudinal modulus (blue line) of the isotropic film laterally restricted near the tricritical transition under the assumption of the homogeneous state. The axes are normalized with respect to $\lambda$ and the temperature of the tricritical transition, $T_{tc}$, $\lambda = 2.5\mu$. The shaded region is the range of temperature where bulk modulus of the film is negative. The horizontal dashed line indicates $K = -\mu/3$ and the temperature at which bulk modulus intersects this dashed line (denoted by the green dashed line) is loss of elastic stability of the homogeneous phase in film with upper and lower free surfaces (obtained from preliminary calculations not shown in the current paper). We do not consider the temperature dependence of the moduli in the HSP region and assume it is constant. LSP: Low symmetry phase, HSP: High symmetry phase.

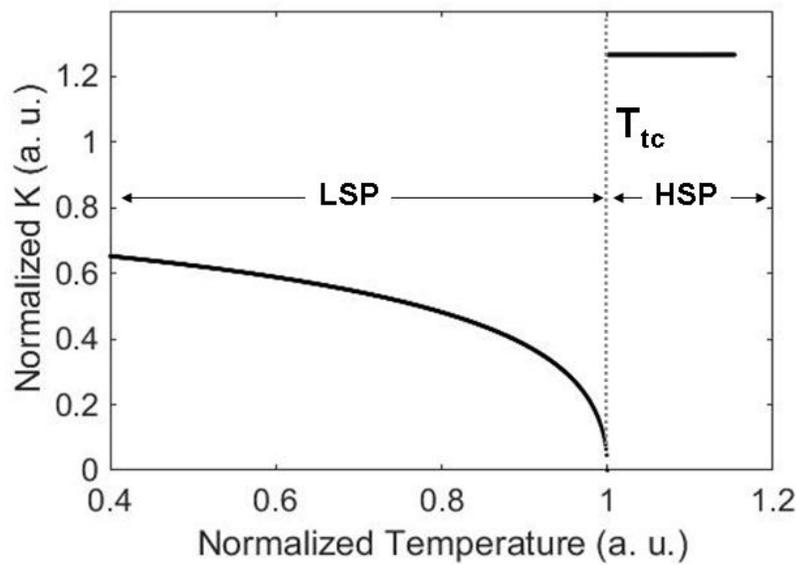

**Figure 1**

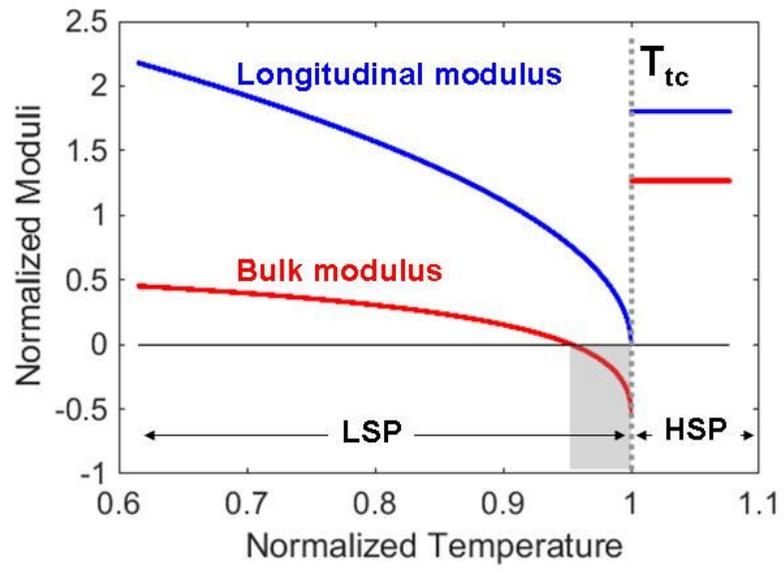

**Figure 2**

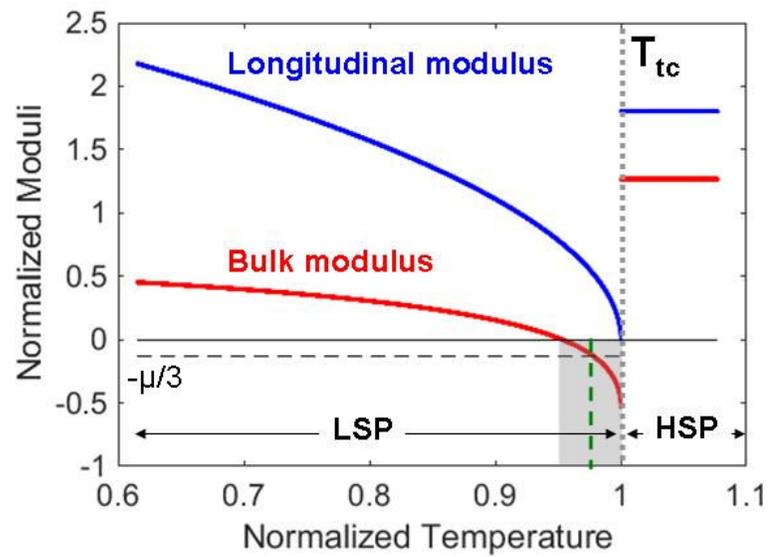

**Figure 3**